\documentclass[%
 aip,
 jmp,%
 amsmath,amssymb,
 reprint,%
]{revtex4-1}

\usepackage[english]{babel}

\usepackage{epsfig}
\usepackage{natbib}
\usepackage{color}
\usepackage{amsmath}
\usepackage{amssymb}

\usepackage{graphicx}
\usepackage{dcolumn}
\usepackage{bm}

\newcommand{\mon}{\begin{displaymath}}
\newcommand{\moff}{\end{displaymath}}

\newcommand{\od}[2]{\frac{d {#1}}{d {#2}}}

\newcommand{\eon}{\begin{equation}}
\newcommand{\eoff}{\end{equation}}
\newcommand{\eaon}{\begin{eqnarray}}
\newcommand{\eaoff}{\end{eqnarray}}

\newcommand{\appropto}{\mathrel{\vcenter{
  \offinterlineskip\halign{\hfil$##$\cr
    \propto\cr\noalign{\kern2pt}\sim\cr\noalign{\kern-2pt}}}}}

\newcommand{\kp}{k_\perp}
\newcommand{\kl}{k_\parallel}
\newcommand{\kt}{k_\theta}

\begin{document}

\preprint{AIP/123-QED}

\title 
{Coupling of alpha channeling to parallel wavenumber upshift in lower hybrid current drive}

\author{I. E. Ochs}
\affiliation{Department of Physics, Harvard University, Cambridge, Massachusetts 02138}

\author{N. Bertelli}
\affiliation{Princeton Plasma Physics Laboratory, Princeton, New Jersey 08543}

\author{N. J. Fisch}
\affiliation{Department of Astrophysical Sciences, Princeton University, Princeton, New Jersey 08540}
\affiliation{Princeton Plasma Physics Laboratory, Princeton, New Jersey 08543}

\date{\today}

\begin{abstract}
Although lower hybrid waves have been shown to be effective in driving plasma current in present-day tokamaks, they are predicted to strongly interact with the energetic $\alpha$ particles born from fusion reactions in eventual tokamak reactors.
However, in the presence of the expected steep $\alpha$ particle birth gradient, this interaction can produce wave amplification rather than wave damping.
Here, we identify the flexibilities and constraints in achieving this amplification effect through a consideration of symmetries in the channeling interaction, in the wave propagation, and in the tokamak field configuration.
Interestingly, for standard LH current drive that supports the poloidal magnetic field, we find that wave amplification through $\alpha$ channeling is fundamentally coupled to the poorly understood $|\kl |$ upshift. 
In so doing, we show that wave launch from the tokamak high-field side is favorable both for $\alpha$-channeling and for achieving the $|\kl |$ upshift.
\end{abstract}

\maketitle


\section{\label{sec:level1}Introduction and Motivation}

Lower hybrid (LH) waves are predicted to be effective in driving substantial plasma current in tokamaks  \cite{fisch1978confining},
an effect that has enjoyed  extensive  demonstration in  tokamak experiments   \cite{fisch87}.
Yet  there remains  a concern that,  in extrapolating to a fusion reactor, high-energy $\alpha$ particles born in the plasma core could strongly damp the LH wave, thus significantly reducing current drive efficiency \cite{wong,fisch_92a,wang2014influence}.
Fortunately, by  coupling diffusion in energy to diffusion in space  (known as \emph{alpha channeling}), 
 a favorable population inversion may appear along the diffusion path,   
causing the $\alpha$ particles to amplify rather than damp the wave \cite{fisch1992interaction}.

Recently, launching  the LH wave from the tokamak high-field side (``inside launch'')  was proposed  to enable the LH wave  to more deeply penetrate the plasma core, with the waveguide better protected from plasma-material interactions \cite{podpaly2012lower,sorbom2014arc}.
Since in a reactor,  $\alpha$ particles would be abundant close to the plasma center,  
the question arises whether  interactions of deeply penetrating waves with $\alpha$ particles can be made favorable, while preserving high current drive efficiency.   
Despite many ray-tracing studies of LH waves to optimize the current drive effect 
\cite{Imbeaux,Decker,Peysson,Ceccuzzi,Horton,Hillairet,Nilsson,Shi,schneider2009self,spada1991absorption,barbato1991quasi,barbato2004absorption,bonoli1987radiofrequency},  
no study has optimized jointly for  LH current drive and $\alpha$-channeling. 

It turns out that symmetries in the LH dispersion relation constrain the possibilities in achieving this joint optimization.
The channeling effect depends  on the sign of poloidal wavenumber $k_\theta$ \cite{fisch1992interaction},
a dependency exploited in ion Bernstein waves \cite{fisch_95a}, where particularly large wavenumbers could be arranged as a result of mode conversion
\cite{valeo1994excitation}. 
For the LH wave, 
$k_\theta$ similarly determines the channeling condition.
However, it is the LH toroidal wavenumber $k_\phi$ that determines the current drive direction through the wave interaction with electrons.
Thus joint optimization of the current drive and $\alpha$-channeling effects requires understanding the joint evolution of $k_\theta$ and $k_\phi$, which are determined by the launch geometry  \cite{bonoli1982toroidal}.  

Through consideration of these symmetries, we find in the analysis presented below that optimization is further constrained by an inescapable coupling of $\alpha$-channeling to the so-called ``$|\kl|$ upshift," where $k_\|$ is the wavenumber parallel to the magnetic field.
This increase in $k_\|$ along the ray trajectory decreases the resonant thermal velocity of the ray, and thus is thought to resolve the so-called ``spectral gap" puzzle in LH wave interactions \cite{bonoli1986simulation}, wherein injected LH waves interact with the plasma despite being injected with a super-resonant parallel phase velocity.
While well documented experimentally 
\cite{bernabei_82,porkolab1984observation,karney_fisch_jobes}, this upshift continues to elude definitive explanation.

By considering a tokamak geometry with circular and concentric flux surfaces, we thus derive fundamental symmetries that constrain the joint optimization of current drive and $\alpha$-channeling.
Interestingly, we find that an upshift must occur for LH waves that both support the channeling effect and drive current supportive of the poloidal magnetic field, which occurs during inside launch.

\section{Channeling direction}
\begin{figure}[t] 
	\center{\includegraphics[width=1\linewidth]{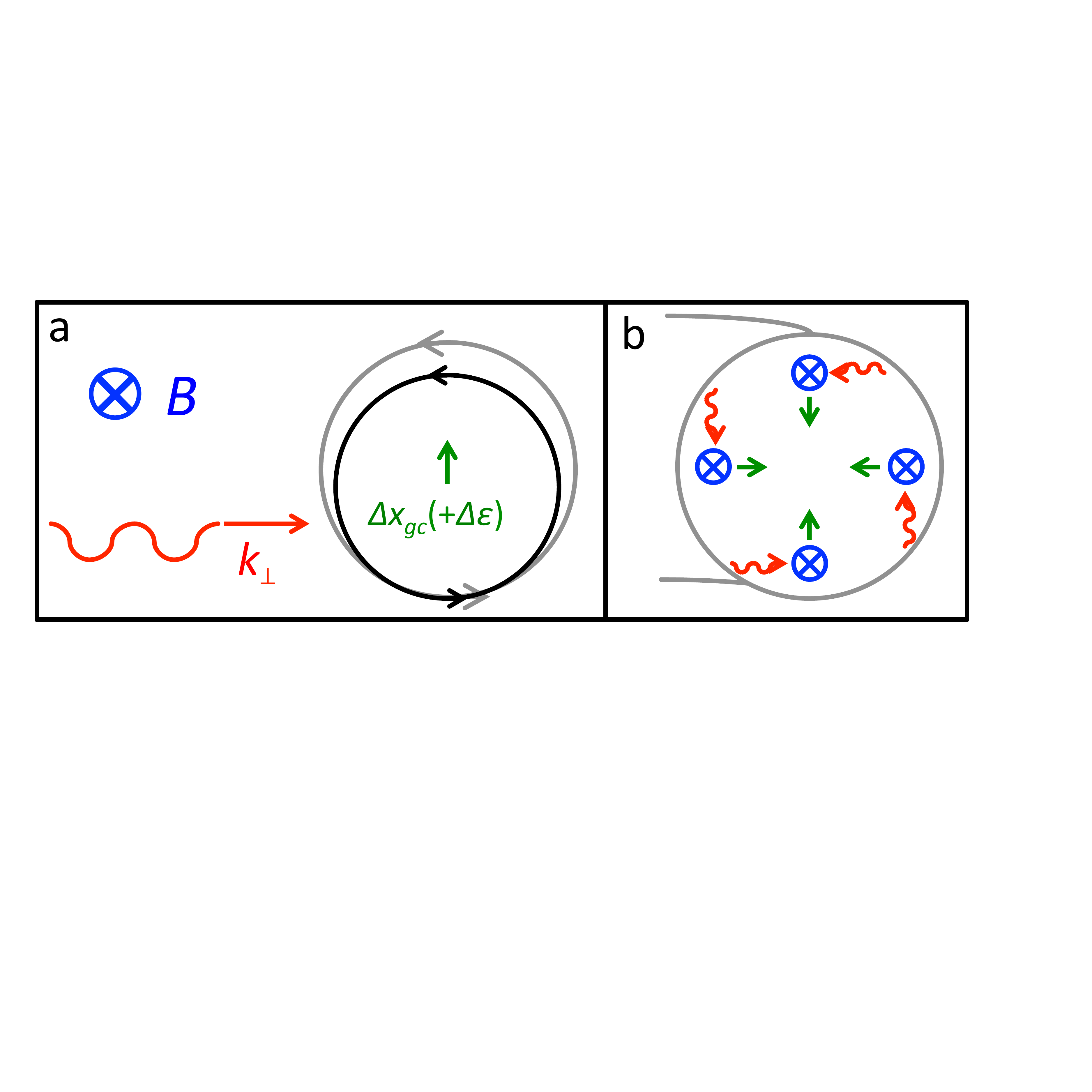}}
	\caption{(a) Schematic of the channeling effect, showing the coupling between the spatial displacement $\Delta x$ and energetic displacement $\Delta \epsilon$ due to interaction with the LH wave \citep{fisch1992interaction}. When wave energy is channeled into the $\alpha$ particles, they move in the direction of $\bf{k} \times B$. (b) In a tokamak magnetic field configuration, $\bf{B}$ (blue) is approximately aligned with the toroidal tangent vector $\hat{\bf{\phi}}$, and so a radially inward pointing $\bf{k}\times B$ (green) requires positive $\kt$ (red). In this case, $\xi \approx - \kt/\kp$. Color available online.}
	\label{fig:xgc}
\end{figure}

To see how  channeling  is constrained under LH wave propagation, consider  that $\alpha$ particles that gain energy from the interaction move in the direction of $\bf{k} \times \bf{B}$, while those that lose energy move in the direction of $-\bf{k} \times \bf{B}$ (Fig. \ref{fig:xgc}a).
Since the interaction tends to be diffusive, on whether the $\alpha$ particles on average gain or lose energy from the wave depends on the distribution of  $\alpha$ particles
along the diffusion path. 
For  interactions with an electrostatic wave in a magnetized homogeneous plasma,  an isotropic distribution of $\alpha$ particles would tend to gain energy, because the projection of the distribution function on any one direction would be monotonically decreasing in energy. 
However, in the presence of a radial $\alpha$-particle gradient, the coupling with spatial diffusion means that $\alpha$ particles can be diffused on average from high energy at the high-density plasma core to low energy at the plasma periphery, thus transferring energy out of the $\alpha$  particles and into the wave.
Because this requires particles that transfer energy \emph{to} the wave to move \emph{outward}, particles that receive energy from the wave must be pushed toward the plasma core, and thus $\bf{k} \times \bf{B}$ must point inward at each flux surface. 
 
Consider therefore a tokamak with concentric, circular magnetic surfaces (Fig. \ref{fig:xgc}b), so that  the flux surface normal vector is given by the minor radius vector $\bf{\hat{r}}$.
Of interest then is the sign and magnitude of 
\begin{equation}
\xi \equiv \frac{\bf{k} \times \bf{B}}{|\bf{k} \times \bf{B}|} \cdot \bf{\hat{r}}.
\end{equation}
The magnitude of $\xi$ represents the extent of the push received by the $\alpha$ particles that occurs in the radial direction, while the sign represents the direction of channeling: when $\xi$ is negative, particles that gain energy will be pushed to the plasma center. 
Thus, $\xi$ must be negative to reduce or reverse the damping.

In a tokamak, the magnitude of the toroidal magnetic field generally greatly exceeds that of the poloidal field, i.e. $|B_\phi| \gg |B_\theta|$.
For lower hybrid waves, we also generally have  $\kp \gg k_\parallel$.
It then follows that
\begin{equation}
	\xi \approx - \frac{\kt}{|\kp |},
\end{equation}
where $|\kp | \approx \sqrt{\kt^2 + k_r^2}$. 
Thus for $B_\phi>0$ (as in Fig. \ref{fig:xgc}b) proper channeling requires $\kt$ to be positive,
and in general we must have  $B_\phi \kt>0$.

For the LH wave, the sign of $\kt$ turns out to be intimately connected with the poloidal position of the ray: specifically, $\kt$ tends to decrease along the ray above the poloidal equator ($0<\theta<\pi$), and increase below it. 
This effect is \emph{independent of both the direction of current drive and the direction of the the poloidal and toroidal magnetic fields}.
To see this independence, consider a simple, well-known electrostatic model of dispersion for $\Omega_{i}^2 \ll \omega^2 \ll \Omega_{e}^2$ in a tokamak of major radius $R_0$ \cite{bonoli1982toroidal}: 
\begin{align}
	D_0 &\approx  \left(c^2/\omega^2\right) \left(\kp^2 -  (\omega_{pe} / \omega)^2 \kl^2\right) \label{eq:disp}\\ 
	B_\phi &= B_{\phi 0}    / [  {1+ ({r}/{R_0}) \cos \theta} ].
\end{align}
Although this model is strictly valid only near $\kt = 0$, simulations show that the symmetries uncovered here also hold for the full electromagnetic cold-plasma dispersion relation.
For tokamaks, $B_\phi \gg B_\theta$ and $\omega_{pe} \gg \omega$, so that the initial evolution of $\kt$ (when $\kt \simeq 0$) is determined by 
\begin{align}
	\od{\kt}{t} &= \frac{\partial D_0/ \partial \theta}{\partial D_0/ \partial \omega} \notag\\
	& \approx - \left(\frac{B_\theta^2}{B_{\phi 0}^2} \right) \left(\frac{\omega^3 (R_0 + r \cos{\theta})}{2 \omega_{pe}^2 R_0^2}\right) \sin{\theta} , \label{eq:ktsym}
\end{align}
 and
\begin{align}
	 \od{\theta}{t} &= -\frac{\partial D_0/ \partial m}{\partial D_0/ \partial \omega} 
         \approx       
          \left(\frac{B_\theta}{B_{\phi 0} k_\phi}\right) \left(\frac{\omega (R_0 + r \cos{\theta})}{2 r R_0}\right). \label{eq:tsym}
\end{align}
For  current drive supporting the poloidal magnetic field ($k_\phi B_\theta > 0$),  
and for proper channeling  ($\kt B_\phi>0$), it follows  that $B_\phi>0$ requires $d\kt/dt>0$. 
Since $d\kt/dt \propto -\sin \theta$, $\kt$ will increase along the ray when $180^\circ<\theta<360^\circ$, and thus the majority of the ray trajectory must occur \emph{below the poloidal equator} to ensure proper channeling in a circular tokamak.

\begin{figure}[b] 
	\center{\includegraphics[width=1\linewidth]{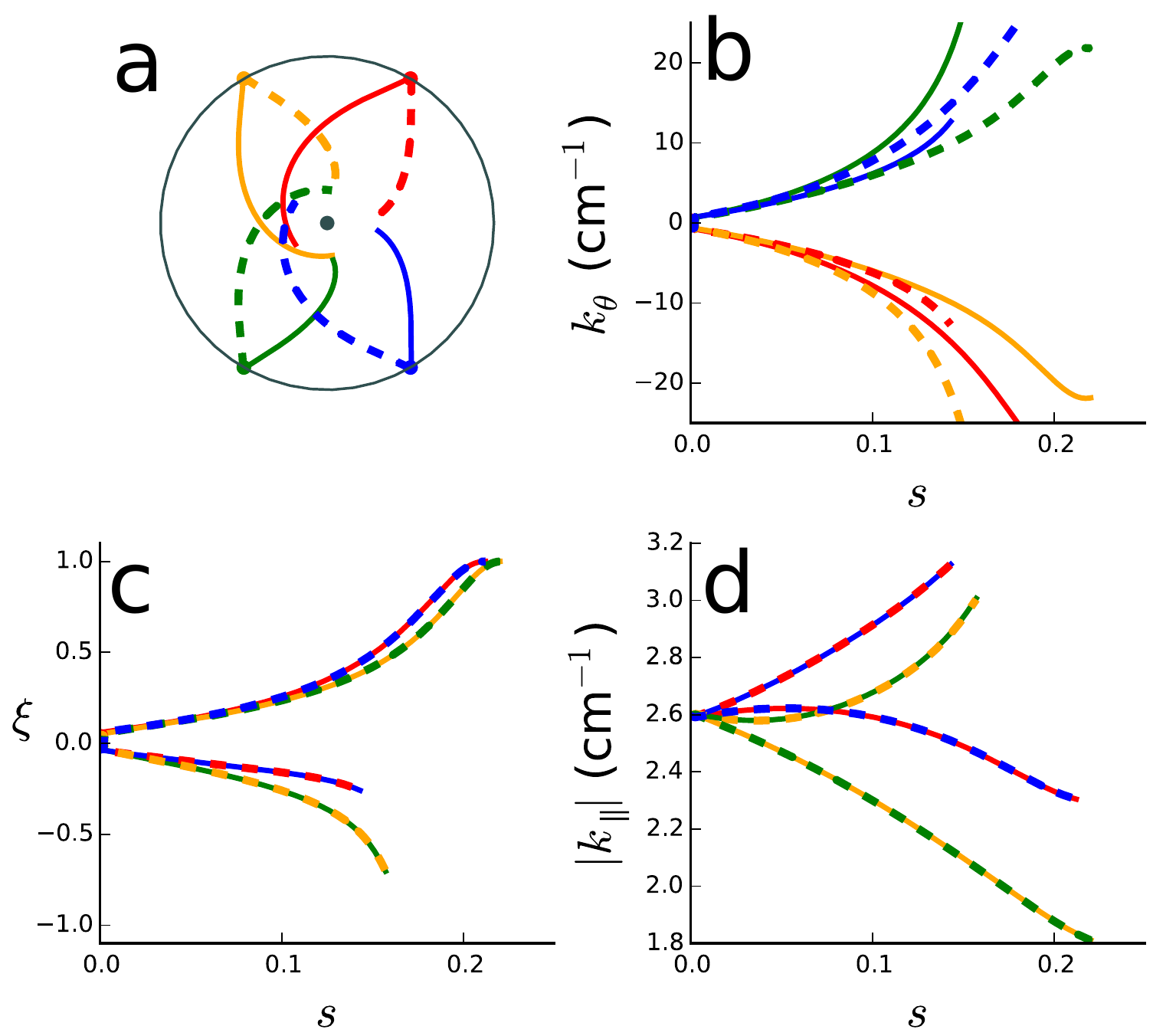}}
	\caption{Ray propagation symmetries with respect to $B_\phi$ reversal in circular tokamak of minor radius $a=$17 cm.
	Core electron density and temperature are $5\times 10^{13}$ cm$^{-3}$ and 5 keV respectively, declining parabolically to $1\times 10^{13}$ cm$^{-3}$ and 500 eV at the plasma periphery.
	Wave parameters were $n_\phi = -2.7$ and $f = 4.5$ GHz.
	Ray trajectories were simulated using geometrical optics code GENRAY \cite{smirnov2001genray}, employing the electromagnetic cold-plasma dispersion relation (in contrast to the simplified electrostatic model in Eq. \ref{eq:disp}).
	Solid lines:  $B_\phi>0$;  dashed lines:  $B_\phi<0$. The poloidal trajectories are shown in (a). Reversing $B_\phi$ and $\theta$ results in antisymmetric evolution in $\theta$ and $\kt$ as a function of poloidal distance $s$ in meters (a-b). However, the reversals have no effect on $\xi$ or the $|\kl|$ upshift, which are clearly coupled (c-d). Color available online.}
	\label{fig:ktsym}
\end{figure}

The key point here is that the $\kt$ near the plasma periphery cannot be too large.  It is basically on the order of the parallel wavenumber, as both will be dictated by the physical dimensions of the launching structure.  However, for lower hybrid waves, the full perpendicular wavenumber is on the order of the $\sqrt{m_i/m_e} \sim 50$ greater than the parallel wavenumber.   
Furthermore, although the radial wavenumber will dominate at the plasma periphery, near the plasma center the perpendicular wavenumber will point essentially in the azimuthal direction.   
Hence, the magnitude of $\kt$  grows from on the order of $k_\parallel$  at the periphery to  substantively higher values.  Because the initial value of $\kt$  is  negligible compared to the final value, it follows that it is the sign of $d\kt/dt$ rather than the initial condition on $\kt$  that plays the critical role.  Hence, several symmetries of the channeling effect may be derived from Eqs.~(\ref{eq:ktsym}) and (\ref{eq:tsym}).   These symmetries are also confirmed by considering the full geometrical optics ray equations as done in Fig.~\ref{fig:ktsym}.

\section{$B_\theta$ reversal and $k_\parallel$ upshift}
For  current drive supportive of the poloidal magnetic field ($k_\phi B_\theta >0$), the sign of $B_\theta$ turns out to have no impact on the poloidal trajectory or the evolution of $\kt$.
This follows directly from the fact that Eqs.~(\ref{eq:ktsym}) and (\ref{eq:tsym}) depend only on $B_\theta^2$ and  $B_\theta/k_\phi$. 
Thus $\alpha$-channeling is unaffected under coupled reversal of $k_\phi$ and  $B_\theta$.

\emph{$B_\phi$ reversal}:
Under $B_\phi$ reversal, proper channeling requires $\kt$ reversal as well. 
Since Eq. (\ref{eq:ktsym}) depends only on $B_{\phi 0}^2$, 
it follows that  $\sin \theta>0$, corresponding to optimized launch from above the poloidal equator. 
Interestingly, this sign reversal in $d\kt/dt$ ensures that, all other quantities equal, a launch with ($\kt$,$\theta$) when $B_\phi>0$ will have a perfectly antisymmetric poloidal trajectory and $\kt$ evolution to the launch with  (-$\kt$,-$\theta$) when $B_\phi<0$, since Eq.~(\ref{eq:tsym}) also changes sign with $B_\phi$ (Figs.~\ref{fig:ktsym}a and \ref{fig:ktsym}b).

Consider now that  the $|\kl |$ shift is determined by the magnitude and sign of $(\kt \cdot B_\theta) (k_\phi \cdot B_\phi) = (\kt B_\phi) (k_\phi B_\theta)$.
Since for supportive current drive $k_\phi B_\theta>0$, and for proper channeling  $\kt B_\phi>0$, launching to ensure  proper channeling gives $\Delta|\kl| > 0$ in the region of strong channeling regardless  of the magnetic field geometry (Fig.~\ref{fig:ktsym}d).

These constraints apply only for supportive current drive.  
However, 
for current drive that opposes $B_\theta$ ($k_\phi B_\theta <0$), for example for current profile control,
Eqs.~(\ref{eq:ktsym}) and (\ref{eq:tsym}) then show that although the sign of $d\kt/dt$ stays the same, the sign of $d\theta/dt$ is reversed, breaking the anti-symmetric evolution  observed under $B_\phi$ reversal. 
Interestingly, the trajectory resulting from reversed current drive for proper channeling is symmetric with respect to the trajectory resulting from supportive current drive for improper channeling ($k_\theta B_\phi <0$).
Thus, since $k_\phi$ reverses sign while all other quantities remain the same, reversed current drive results inevitably in a $|\kl|$ downshift in the region of strong channeling. 

These  symmetries and their associated constraints, imposed  through the fundamental coupling in LH current drive of $\alpha$ channeling to the $|\kl|$-upshift, are our the key results. 
Although derived strictly for concentric circular flux surfaces, they are valid in a more general magnetic geometry in the regime of interest, namely where the LH waves penetrate deeply enough to encounter $\alpha$ particles
at the locally circular plasma center.

\section{Discussion}
When $B_\phi>0$ and $k_\phi$ supports the plasma current, a ray trajectory that spends most of its approach to the region of high $\alpha$-particle density below the poloidal equator is optimal for wave amplification, since its poloidal wavenumber becomes both large (on the order of $k_\perp$) and of the right sign.
When the tokamak is large enough that the ray sweeps a significant range of $\theta$, this generally occurs for wave launch from the high-field side.  
Thus we can see that LH waves launched from the tokamak's high-field side, which are already predicted to be advantageous from an engineering standpoint  \cite{podpaly2012lower,sorbom2014arc}, can also avoid $\alpha$-particle damping or even experience amplification as they penetrate near the plasma center.  
For concentric circular flux surfaces, we derived that the wave amplification conditions also necessarily result in a $|\kl|$ upshift in the region of strong damping for waves used to support the plasma current, and a $|\kl|$ downshift for those used to counteract the plasma current.
For the general tokamak, this property will apply near the magnetic axis, where the geometry is effectively circular and where the key interactions take place.  

Here we have focused on amplification of the wave energy used for current drive.  
After all, it was the damping of the lower hybrid wave on electrons in a homogeneous plasma \cite{wong} that instigated the search for the channeling effect in the first place.  
However, it should be noted that channeling leads to other potential benefits.  
For example, the same waves that diffuse $\alpha$ particles from high energy at the center to low energy at the periphery also have the potential to diffuse cold fuel ions from the periphery (where they are dense) to hot in the center (where the density of very hot fuel ions is small).  
When damping on these cold ions dominates, the lower hybrid wave would  transfer energy from  outward-moving $\alpha$ particles to inward-moving fuel ions.  
Such an effect could lead to a hot ion mode, which is significantly advantageous for fusion \cite{fisch_94}.  

Of course, both the current drive and ion heating effects could be present simultaneously.
After all, waves that channel energy out of the $\alpha$  particles while ejecting them will have the right sign of poloidal mode number to heat fuel ions while pinching them to the center.  
Thus the potential benefit of channeling energy from $\alpha$ particles is large.

Given the upside potential offered by $\alpha$ channeling, it is important to test whether the lower hybrid wave can stably absorb energy from the $\alpha$ particles.  
The easiest differential test would be to leave all other parameters equal, but just to reverse $k_\theta$, with one sign resulting in damping and the other in amplification.  
Unfortunately, as derived here, the strict coupling of $k_\theta$ to the $|\kl|$ upshift renders this experiment impossible, since the upshift itself dramatically impacts the driven current in several ways.
Most directly, it reduces the electron-resonant parallel velocity, decreasing current drive efficiency.
Less directly, it also could cause the damping of the wave to take place in an entirely different location of the plasma, possibly near the periphery at lower density, which might then cause the current drive efficiency to increase. 
Either way, it makes it impossible to perform a differential test simply by reversing $k_\theta$.

In light of this coupling, the best differential test might instead be to arrange for wave conditions suitable for the channeling effect, both in the presence and absence of an energetic (MeV) beam of ions (such as might occur under minority ion cyclotron heating).  
Suppose that the energetic ions are arranged to be mainly in the central region of the plasma, with sharp spatial gradients.  
The presence of the ions should then amplify the lower hybrid wave for one sign of $\kt$, and damp it for the opposite sign, leading to a detectable change in the driven current.
Similarly, with $k_\phi$ reversed, the presence of the ions could increase or decrease the counter-directed current, which would be measurable as well.
Thus, a multiplicity of comparisons could indicate the effect.

Note that a similar differential test using neutral beams was performed successfully to test channeling effects predicted for the ion Bernstein wave \cite{fisch_IAEA,clark_00}.  
However, this test employed relatively low-energy neutral beams, so it was not possible to launch waves that would retrieve the ion energy.
Instead, waves were launched that would heat the ions along predicted diffusion paths, with the main measured effect being the movement of the ions along these paths.
In the case of lower hybrid waves, the diffusion paths may not extend all the way to the periphery, necessitating a measurement based on internal effects such as current drive.

\section{Conclusion}
Here we have shown that achieving favorable $\alpha$ channeling while driving plasma current supportive of the poloidal magnetic field necessitates two main effects.
First, it imposes a fundamental preference for high-field-side LH launch, adding to the potential advantages of this newly-proposed launcher configuration.
Second, it leads necessarily to a $|\kl|$ upshift, constraining the possible experimental tests of the channeling effect.
The identification of these strong constraints was derived theoretically and simulated for a circular cross-section tokamak.  
Nevertheless, it remains to explore in greater detail the joint optimization of current drive and  $\alpha$ channeling in specific tokamak geometries.

 {  \color{blue}  

}

\begin{acknowledgments}
The authors would like to thank P. Bonoli for useful discussions. This work was performed under U.S. DOE contract DE-AC02-09CH11466. One of us (IEO) thanks the support of the National Undergraduate Fellowship Program in Plasma Physics and Fusion Energy Sciences.

\end{acknowledgments}


\bibliography{ref4_njf,more_coupling}

\clearpage
\newpage

\end{document}